\documentclass[final,onefignum,onetabnum]{siamonline190516}

\usepackage{lipsum}
\usepackage{amsfonts}
\usepackage{graphicx}
\usepackage{epstopdf}
\usepackage{algorithmic}
\ifpdf
  \DeclareGraphicsExtensions{.eps,.pdf,.png,.jpg}
\else
  \DeclareGraphicsExtensions{.eps}
\fi

\usepackage{enumitem}
\setlist[enumerate]{leftmargin=.5in}
\setlist[itemize]{leftmargin=.5in}

\usepackage{caption}
\usepackage{subcaption}
\usepackage{bbm}

\newcommand{\R}{{\mathbb{R}}}

\newsiamremark{remark}{Remark}
\newsiamremark{hypothesis}{Hypothesis}
\crefname{hypothesis}{Hypothesis}{Hypotheses}
\newsiamthm{claim}{Claim}

\headers{Linear Run Time of Persistent Homology Computation with GPU Parallelization}{Michael G. Rawson}

\title{Linear Run Time of Persistent Homology Computation with GPU Parallelization
}

\author{Michael G. Rawson\thanks{Department of Mathematics, University of Maryland at College Park, Maryland, USA 
  (\email{rawson@umd.edu}).}}

\usepackage{amsopn}

\begin{document}

\maketitle

\begin{abstract}
Persistent homology is a crucial invariant that is used in many areas to understand data. The $O(N^4)$ run time is a hindrance to its use on most large datasets. We give a parallelization method to utilize multi-core machines and clusters. We implement the computation of the $0^{th}$ persistent homology with OpenMP parallelization and observe a 1.75 fold performance increase by using 2 threads on a dual core machine. We also benchmark the computation using larger numbers of threads and show that the thread computational overhead decreases performance. With GPU parallelization, we analytically and empirically decrease the run time scaling from $O(N^4)$ to $O(N^3)$ and even $O(N^2)$ where $N$ is the number of data points, for a large enough GPU. Next, we analytically show run time scaling $O(N)$ for an even larger GPU.
\end{abstract}

  



\section{Introduction}

Topology originated in the study of holes. Topology is now defined as the collection of open sets associated to a space. By treating the holes of a space as a basis of a module over a ring (or vector space over a field), we can work with simple objects that represent complicated topological spaces. After defining this carefully, we call this the homology, see \cite{Giusti2016, Purvine2018}. What also comes out is that there are holes of different dimension. Essentially, the $k^{th}$ homology of a space is the vector space on the basis of $k^{th}$ dimensional holes. In applications, we often care about clustering data or detecting bifurcations (or branching) in data. The $0^{th}$ homology describes the cluster structure. This also corresponds to the bifurcation structure when computed locally. However, given mere data points, the topology is a trivial discrete topology. We need to make some assumptions to get a meaningful topology of the underlying space of the data. 

Let us introduce the Vietoris-Ripps ($VR_\epsilon$) complex from \cite{hausmannvietoris}. The $VR_\epsilon$ complex takes a small positive number $\epsilon\in\R$ and assumes that data points with pairwise distance less than $\epsilon$ are connected by an edge, creating a graph. A graph is a dimension 1 simplicial complex. We will only consider dimension 1 $VR_\epsilon$ complexes because higher dimensions don’t affect the $0^{th}$ homology. If one desires to compute higher dimension $VR_\epsilon$ complexes, that is simply done by applying the recursive rule that there will be a $k$-simplex on some $k$ points if and only if every $j$-subset of these points has a $j$-simplex for every $k>j>1$. What is missing here is what $\epsilon$ to use. The resulting homology drastically changes as $\epsilon$ is changed. By essentially plotting the homology over $\epsilon$ then we can get an idea of the best $\epsilon$. To define this carefully, instead of a plot we’ll define intervals corresponding to the basis elements of the homology. The starting value of each interval will be the smallest $\epsilon$ that results in the corresponding homology basis element. The ending value of each interval will be the largest $\epsilon$ that results in the corresponding homology basis element. When $\epsilon$ is near 0, the $VR_\epsilon$ graph has a vertex for each point but no edges. As $\epsilon$ increases, many edges are added based on how the data points are distributed. Then as $\epsilon$ increases more, the edge additions tapers off. The edge additions will remove basis elements from the homology when the edge connects two disconnected subgraphs. With these assumptions, there will be many short intervals and few long intervals. The long intervals will correspond to the topology of the space. These intervals, called ``barcodes", are the persistent homology of the space. There is an algorithm to compute the persistent homology or barcodes, which we describe next.

\section{Barcode Algorithm}

We explain the barcode algorithm for the $0^{th}$ persistent homology, see \cite{carlsson_persistence_2005, 7472911, ghrist_barcodes_2007}. We start with a list of $N$ data points, $X$, each in $d$ dimensional Euclidean space, $E^d$. Then we calculate the distance between every pair of points, which corresponds to the $\epsilon$ at which an edge is added between the pair in $VR_\epsilon$. Take the list of distances, $E$, and sort it in increasing order. Remove duplicate elements in $E$ and call it $D$. We build a matrix, $M$, with a column for each edge and a row for each vertex in $VR_\infty$ (the complete graph). Set $M(i,j)$ to $t$ a when $i$ is a vertex of edge $j$ and where $a$ is the index of edge $j$ in $D$ which ranges from 1 to the length of $D$. All other entries of $M$ are set to 0. We make the entries to be polynomials of variable $t$ with coefficients restricted to 0 or 1. The next step is to column reduce this matrix so that what remains is lower triangular. Then the remaining nonzero diagonal entries, say $t^b$, correspond to the barcode or interval $(0,b)$. With this list of intervals, the algorithm is complete.

\section{Multi-Core CPU Parallelization}

Parallelization is highly desirable for this algorithm. The computational complexity of this algorithms is first $O(N^2)$ for the pairwise distance computation, actually over a factor of 2 due to the distance symmetry. Then plus $O(N^2 \log(N^2))$ for sorting the edges. Then plus $O(N^2)$ to
compute matrix $M$. Then plus $O(N \cdot N^2 \cdot N)=O(N^4)$ to reduce, with pivoting, the $N$ by $N^2$ matrix $M$. Then plus $O(N)$ to collect the barcode. Altogether this is $O(N^4)$. Even if $N$ is small say $N=1000$ data points, the time requirement is approximately $10^{12}$ in time units proportional to the central processing unit (CPU) clock rate, a huge amount of time. Parallelization can leverage multiple processors together to speed up codes and algorithms. We use OpenMP parallelization in C++ to experiment with improving the performance of the barcode algorithm. To use OpenMP we add ``for" pragmas before the for loops. In the matrix reduction code, the outer loop is not clearly parallelizable so we put the pragma inside the outer loop where there are parallelizable for loops. 

\subsection{CPU Results}

\begin{figure}[h]
\centering
\includegraphics[width=.8\linewidth]{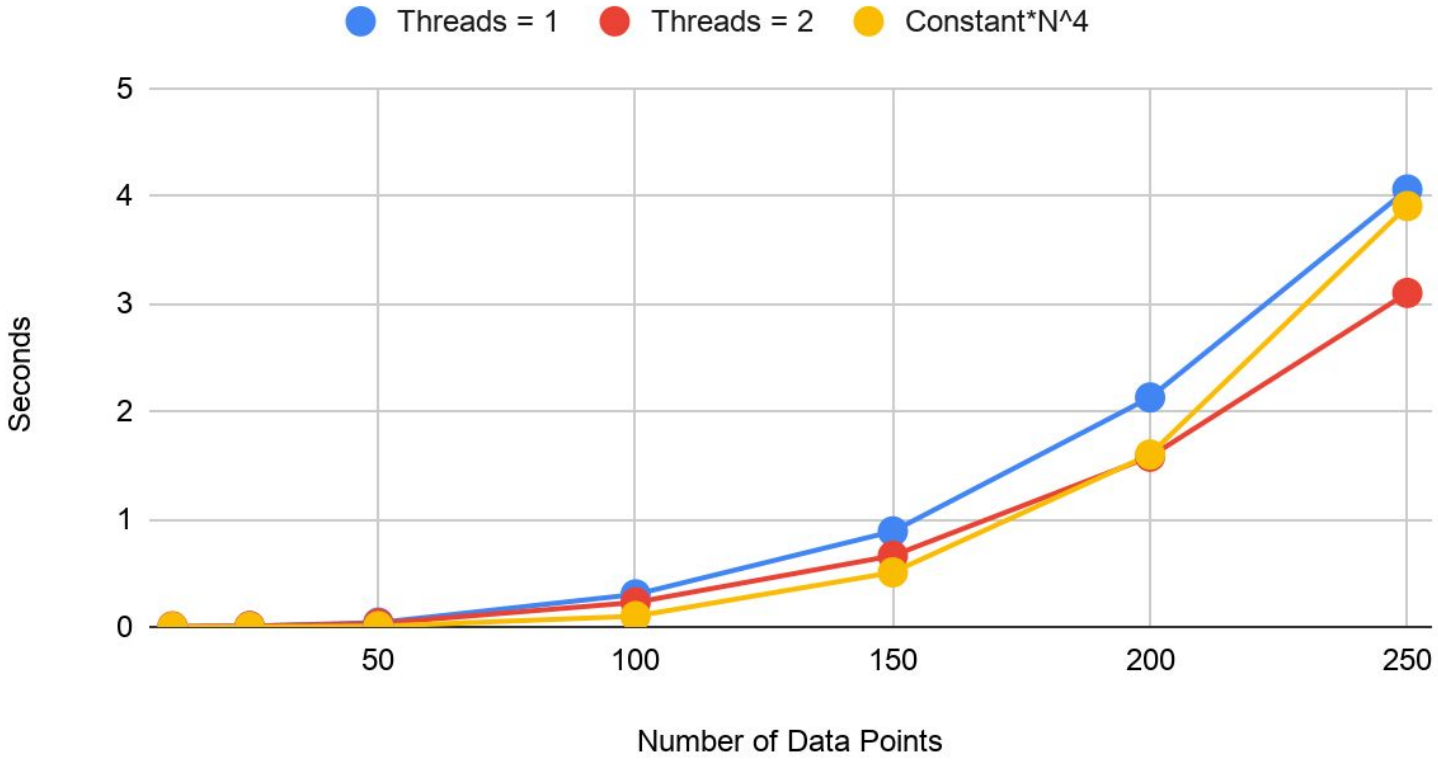}
\caption{Run time to compute the $0^{th}$ persistent homology versus number of data points using 1 or 2 threads and compared with polynomial growth.}
\label{fig:p_hom_par1}
\end{figure}

We implement the algorithm in C++ with OpenMP on a 2 core machine running Ubuntu. We use randomly uniformly distributed pairs of numbers in (0,1) for our data. We compile with g++ with -o3 highest compiler optimization since we want to see if the best performance can be improved with parallelization. We note that with -o0 lowest compiler optimization, we get very similar results. We plot averages of 10 runs. In the best case, the run time would be divided by 2 when going from 1 thread to 2 threads. We plot the experiment in \cref{fig:p_hom_par1}.

\begin{figure}[h]
\centering
\includegraphics[width=.8\linewidth]{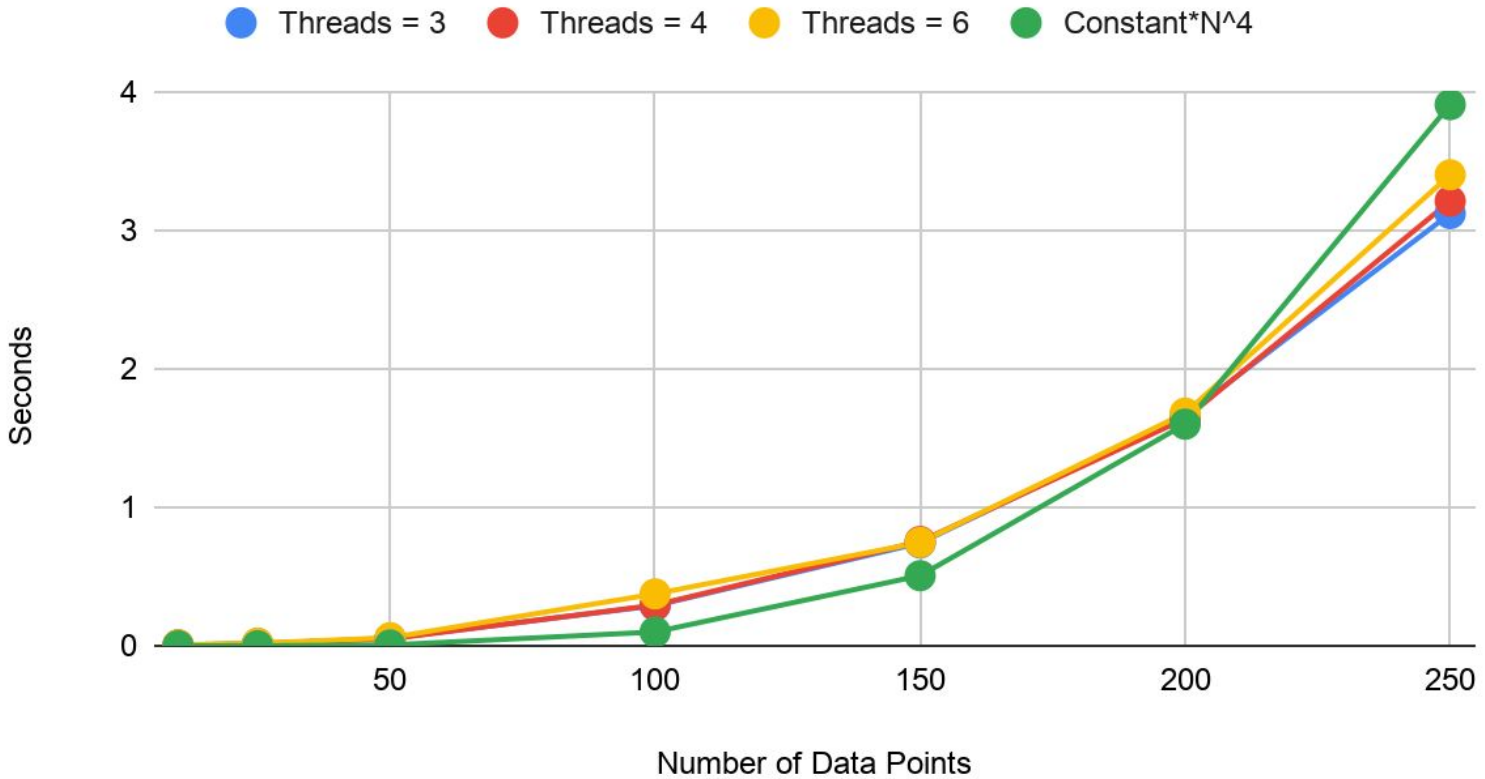}
\caption{Run time to compute the $0^{th}$ persistent homology versus number of data points using 3, 4, or 6 threads and compared with polynomial growth.}
\label{fig:p_hom_par2}
\end{figure}

We get an increasing performance gain with the number of data points. The performance gain is less than double however because the code is not perfectly parallel and due to the thread overhead cost. As for how the run time increases with the number of data points, $N$, we expect $O(N^4)$ from our calculation above. Even with parallelization and dividing the time by 2, we still expect $O(N^4)$. We plot $N^4$ times a constant above and see that our experiments are not far off from $N^4$ though they only need converge as N approaches infinity. Since we use a 2 core machine, more than 2 threads should not give a performance increase. We compare run times with the number of threads greater than 2. We plot the experiment in \cref{fig:p_hom_par2}. We see that more threads beyond the number of processing units decreases performance. This is because threads have an overhead cost. Since a threading point is inside of a loop in the matrix reduction code, the thread overhead cost is multiplied by the number of data points, becoming significant.

We have shown a large performance increase computing barcodes by using parallelization on a dual core machine. We got up to about 1.75 fold performance increase which may approach the limit of 2 fold increase for a large enough number of data points. We also showed that the thread overhead cost can accumulate, destroying performance gains but dependent on the implementation. Repeating this work for the higher persistence homology computations won’t affect much other than increasing the amount of computation.

\section{GPU Parallelization}
First, we use GPU (graphics processing unit) parallelization to compute the $0^{th}$ Persistent Homology. By mass parallelization, we analytically and empirically decrease the run time scaling from $O(N^4)$ to $O(N3)$ and even $O(N^2)$ where $N$ is the number of data points, for a large enough GPU. Next, we analytically show run time scaling $O(N)$ for an even larger GPU.

GPU usage for data analysis has become possible and popular over the past decade. For example, \cite{Murty, Suzuki2012, Suzuki2016} use GPUs for topological data analysis. As these papers show, putting the GPU’s thousands of cores to use in parallel can yield higher performance, that is, lower run times compared to a CPU. As a comparison, consider that an Intel Core i7 980 XE can compute 109 gigaFLOPS \cite{Williams} (floating point operations) while an NVIDIA Tesla P100 can compute 10 teraFLOPS \cite{Harris}. This is achievable because even though the GPU’s clock rate is only about a third of the CPU’s, the GPU has thousands of cores compared to the above CPU’s 6 cores. To get an idea of the parallelization of this GPU, we can compute the ratio of FLOPS over clock rate, so 10 teraFLOPS / 1 Ghz = $10^3$ operations in parallel \cite{Harris}. This compares to the above CPU’s 109 gigaFLOPS / 3 Ghz = 36 operations in parallel which makes sense given about 6 arithmetic logic units (ALUs) per core.

As explained above, the computational complexity of computing the $0^{th}$ persistent homology is $O(N^4)$ where $N$ is the number of data points. This is caused by reducing the boundary matrix of size $N \times N(N-1)/2$. Now, as the matrix reduction iterates down the matrix diagonal, each step is easily parallelizable in constant time, ignoring pivoting. This reduction is then a $O(N)$ operation. So can we experimentally observe total run time growth of $O(N)$?

We use a Red Hat Linux server with an NVIDIA Tesla P100 16 GB GPU. We generate $N$ points of uniformly random data on $[0,1] \times [0,1]$ and then we compute the $0^{th}$ persistent homology described above. We use the GPU to compute the steps of the algorithm in parallel, that is, in constant time. Steps:
\begin{enumerate}
    \item Compute N(N-1)/2 pairwise distances, that is the edges.

    \item Sort the edges.

    \item Create the boundary matrix.

    \item For each diagonal entry of the matrix: Reduction step.

    \item For each diagonal entry of the matrix: Get the barcode interval.
\end{enumerate}

\subsection{GPU Results}

\begin{figure}[h]
\centering
\includegraphics[width=.8\linewidth]{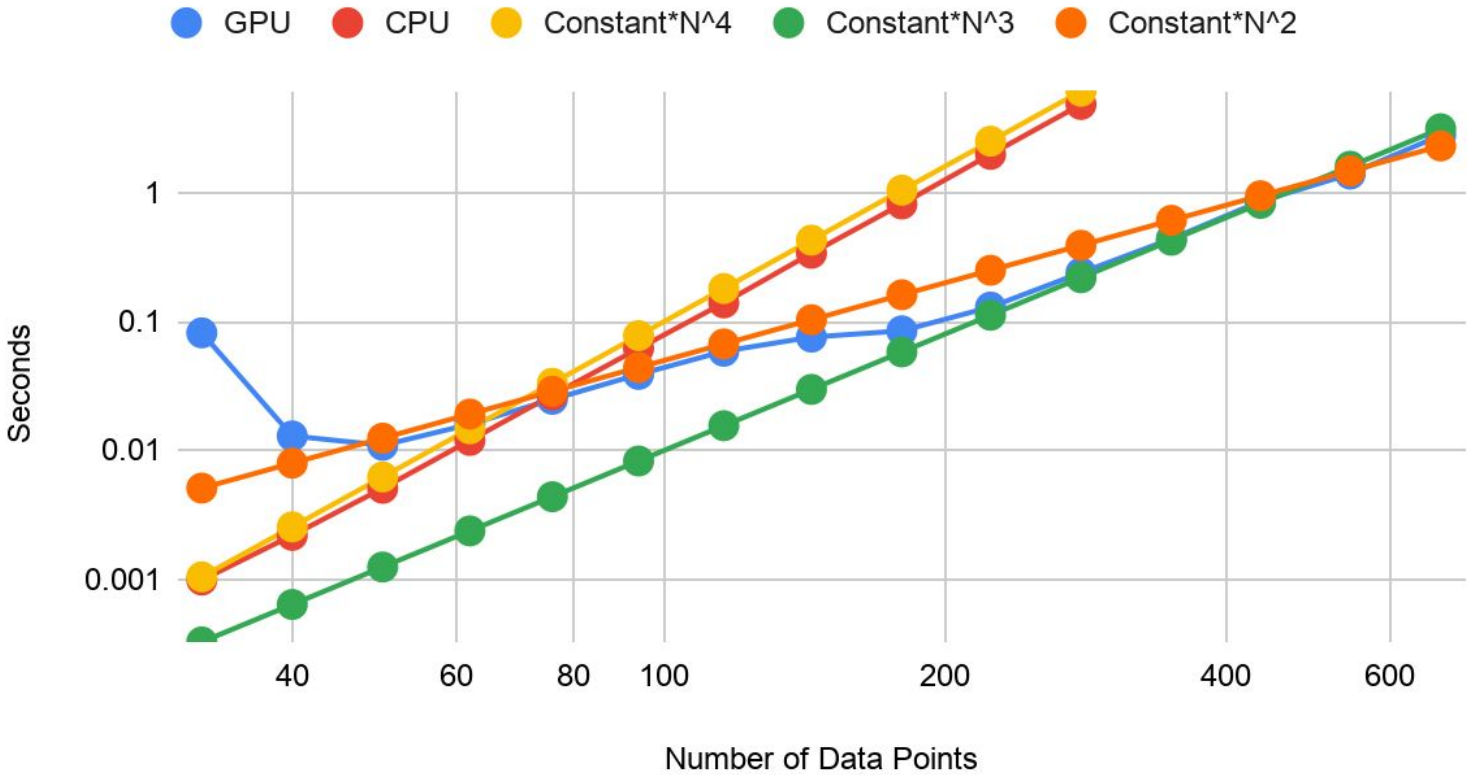}
\caption{Run time to compute the $0^{th}$ persistent homology versus number of data points using CPU or GPU and compared with polynomial growth.}
\label{fig:p_hom_par3}
\end{figure}

Then we plot the run time using the GPU vs sequential CPU code without the GPU, see \cref{fig:p_hom_par3}. The experiments are averages of 10 runs. We notice that the run time of the CPU (no GPU) code grows with $O(N^4)$ as the theory predicts. The run time of the GPU code grows with $O(N^2)$ from $N$=50 to 150. Then the run time of the GPU code grows with $O(N^3)$ from $N$=200 to 700. A note on pivoting, pivoting, to the extent that it is happening, may be thought to be affecting the theory and experiments. Experimentally, even when pivoting is disabled, the result is very similar. As for the theory, finding the pivots could be done with the GPU in logarithmic time by doing a binary search via sums of row slices. Another note on sorting the N(N-1)/2 edges by length, sorting on the GPU in parallel, for a large enough GPU, can be done in $O(\log(N(N-1)/2))$ run time \cite{Powers91parallelizedquicksort}. Our implementation uses the Thrust library for sorting on the GPU and we observe it takes negligible run time.

The reason that we didn’t get $O(N)$ run time growth is because we made a crucial assumption above. We assumed that the GPU could do $N \times N(N-1)/2$ operations in parallel. Using our approximate calculation above of $10^3$, that means $N$ $<$ 28. At those small $N$ values, the memory and initialization times can swamp the calculation. However, if $N$ $<$ $10^3$ or if $N(N-1)/2$ $<$ $10^3$ then we can do column or row, respectively, operations in parallel and then decrease the run time to $O(N^3)$ or $O(N^2)$, respectively. We calculate those thresholds and see $N=150$ and $N=10^3$. This matches the data where we see $O(N^2)$ run time growth up to $N$=150 and then after we see $O(N^3)$ run time growth. We did not run $N$ beyond 700 due to limited time. Of course for $N$ large enough the run time growth must be that of the computational complexity which is $O(N^4)$. Technically, if the computing resources are finite and constant, then they will not change the computational complexity as $N$ grows very large.

\subsection{Conclusions}
We have shown analytically and experimentally that with a large enough GPU, we can decrease the run time growth of computing the $0^{th}$ persistent homology from $O(N^4)$ to $O(N^3)$ and even $O(N^2)$. Analytically, with many large GPUs, $O(N)$ run time growth is possible. However, we’ve ignored memory transfers and latencies which could be researched in future work. Future work also includes the straight forward extension to the higher order homology groups.

\section*{Acknowledgments} 
The author acknowledges the Department of Mathematics and \\ CSCAMM at the University of Maryland at College Park for providing computational resources and support.

\newpage

\bibliographystyle{siamplain}
\bibliography{main}


\end{document}